\documentstyle[aps,preprint]{revtex}
\input epsf
\newcommand{\journal}[4]{{{\sl #1}} {\bf #2}, {#3} (#4)}
\newcommand{\mprb}[3]{\journal{Phys.~Rev.~B}{#1}{#2}{#3}}
\newcommand{\mpre}[3]{\journal{Phys.~Rev.~E}{#1}{#2}{#3}}
\newcommand{\mpra}[3]{\journal{Phys.~Rev.~A}{#1}{#2}{#3}}
\newcommand{\mprl}[3]{\journal{Phys.~Rev.~Lett.~}{#1}{#2}{#3}}

\begin{document}
\draft
%\twocolumn
\title{Slow dynamics and aging in a non-randomly frustrated spin system}

\author{Hui Yin and Bulbul Chakraborty}

\address{Martin Fisher School of Physics, Brandeis University, Waltham, MA
02254.}

\date{\today}

\maketitle

\begin{abstract}

A simple, non-disordered spin model has been studied in an effort to
understand the origin of the precipitous slowing down of dynamics
observed in supercooled liquids approaching the glass transition. A
combination of Monte Carlo simulations and exact calculations indicates
that this model exhibits an entropy vanishing transition accompanied
by a rapid divergence of time scales.  Measurements of various
correlation functions show that the system displays a hierarchy of
time scales associated with different degrees of freedom.  Extended
structures, arising from the frustration in the system, are identified
as the source of the slow dynamics. In the simulations, the system
falls out of equilibrium at a temperature $T_{g}$ higher than the
entropy-vanishing transition temperature and the dynamics below
$T_{g}$ exhibits aging as  distinct from coarsening. The cooling
rate dependence of the energy is also consistent with the usual glass
formation scenario.

\end{abstract}

\section{Introduction}
The exact nature of the glass transition in supercooled liquids is
still an enigma. The hallmark of this transition is a precipitous
slowing down of the dynamics without any accompanying, obvious,
structural changes\cite{Glassrev,Ediger}. The analogous transition in
spin glasses, which have quenched randomness, is much better
understood and the theoretical understanding in that field has
benefitted from the existence of simple microscopic
models\cite{Spinglassrev}.  A search for models which do not have
disorder at the microscopic level but display glassy dynamics has been
one of the important directions of current research.
In this paper, we present a detailed analysis of a non-disordered spin
model whose phenomenology is remarkably similar to that of supercooled
liquids.  The model can be solved exactly in a certain
limit\cite{HuiPRL}, and the exact results can be related to the glassy
dynamics observed in simulations of the model. This connection can be
exploited to address the issue of the origin of the rapid divergence
of time scales\cite{Angell} in supercooled liquids.

It is well known that relaxation times diverge at a critical
point\cite{goldenfeld}, however, unlike critical points, there is no
obvious length-scale divergence that has been associated with the
glass transition. Recent experiments\cite{Ediger,Weitz,Israeloff} and
simulations\cite{Glotzer} point toward the existence of dynamical
heterogeneities\cite{Ediger,Weitz,Israeloff} and a time-dependent
length scale\cite{Glotzer} which grows as the glass transition is
approached.  This raises the question of whether the occurrence of a
critical point is responsible for the anomalously slow relaxations
observed in glasses, and if so what is the nature of this critical
point. A scenario proposed by Adam,Gibbs and DiMarzio is that glassy
dynamics is associated with an underlying configurational entropy
vanishing transition
\cite{Adam,Mezard}. The slow relaxation is due to the small equilibrium entropy
near glass transition temperature and the growing length-scale is
associated with cooperatively relaxing regions. This theory is
compatible with Kauzmann's entropy crisis argument\cite{Kauzmann},
which predicts a lower limit of the glass transition temperature and
is supported by experiments measuring configurational entropy in the
glass forming liquids\cite{Richert}. In recent years, this scenario
has been explicitly realized in p-spin spin glasses as a random first
order transition\cite{Gross,Kirkpatrick,Bouchaud,Wolynes}. Based on an
inherent structures approach\cite{Stillinger}, indications of a similar
scenario have been found in simulations of Lennard-Jones
systems\cite{Sastry}.  Another possibility that has been suggested to
reconcile the observed time-scale divergence with no apparent length
scale divergence is the existence of a zero-temperature critical
point\cite{Sethna} which has characteristics similar to the critical
point in a random-field Ising model\cite{DFisher}.  A major obstacle
in obtaining definitive answers is the identification of  simple,
non-disordered models where such scenarios are explicitly realized.

In recent years, studies have focused on microscopic models with no
quenched disorder which exhibit glassy dynamics.  Models in this
category include a ferromagnetic four-spin plaquette model on a cubic
lattice\cite{Lipowski} which exhibits aging\cite{Swift},
three-dimensional Ising models with competing nearest and next-nearest
neighbor interactions\cite{Shore} exhibiting logarithmic growth.  Besides these, there are constrained dynamics models, as for example, 
the two-dimensional ferromagnetic Ising model with three-spin
interactions \cite{Newman}. The presence of a critical point has not
been related to the glassy dynamics in any of these models.

The model studied in this paper is the compressible,
triangular-lattice, Ising antiferromagnet(CTIAFM)\cite{Lei1}. We will
present results of a Monte Carlo(MC) simulation study of the dynamics
in the supercooled state of CTIAFM and an exact solution in the limit
of vanishing ``thermal'' fluctuations. The exact solution exhibits a
critical point which is
accompanied by a vanishing of configurational entropy.  The model and
its equilibrium behavior are discussed in section II. In section III,
we introduce a mapping of the spin model to a geometrical model and
the exact solution is discussed in section IV.  In section V and VI,
we present our MC study of the supercooled phase and the glass phase,
respectively.  Our conclusions are presented in section VII.

\section{CTIAFM model}

The CTIAFM is a simple extension of the well known triangular-lattice
Ising antiferromagnet(TIAFM)\cite{wannier,houtappel}. The TIAFM is a
fully frustrated model and has no finite temperature transition. The
number of ground states is exponentially large, and the $T=0$ state is
critical, characterized by a power law decay of spin correlation
functions\cite{Stephenson}. The critical ground state has been studied
extensively by mapping to interface models\cite{Blote,Henley}.  The
effect of degeneracy-lifting fields on the TIAFM and the resulting
possibility of phase transitions has also been the subject of many
investigations\cite{houtappel,Blote,Hillhorst,Dhar}. 

In CTIAFM, a
coupling to elastic strain fields removes the degeneracy of the
ground state\cite{Kardar}. In our current study only homogeneous strain
fields are considered; the distortion is uniform and does not depend
on the site.  Thus, the additional degrees of freedom are encapsulated
in three strain fields, \{${e_\alpha, \alpha=1..3}$\}, in the
Hamiltonian:
\begin{equation}
H = J{\sum}_{<ij>}S_i S_j -\epsilon J
{\sum}_{\alpha}e_{\alpha}{\sum}_{<ij>_{\alpha}}S_i S_j + N{E \over
2}{\sum}_{\alpha}e_{\alpha}^{2}~.
\label{model}
\end{equation}
Here $J$, the strength of the nearest-neighbor anti-ferromagnetic
coupling, is modulated by the presence of the second term which
defines a coupling between the spins and the strain fields
\{$e_\alpha$\}. The coupling $\epsilon$ is chosen to be a positive constant to
ensure that antiferromagnetic interaction gets stronger when spins get
closer. The last term represents  the elastic energy needed to stabilize the
unstrained lattice and  $N$ is the total number
of lattice spins. This model can be thought of as a spin model with
anisotropic couplings. The anisotropic couplings are determined by the
strain fields and are, therefore, annealed variables as opposed to
models with fixed (quenched) anisotropy\cite{houtappel,Blote}.

The phase transitions of this model were analyzed in the context of
the effects of elastic strain fields on ordering transition in
alloys\cite{Lei1}. This, and other previous studies, showed that the
competition between energy and entropy leads to a first order
transition at a finite temperature $T_1$\cite{Kardar,Leigu}. Above
$T_1$, the system is in a disordered, paramagnetic state and the lattice
is contracted isotropically in three directions. Below $T_1$, there is
a lattice distortion and the spins order as stripes with alternating
up and down rows: the ferromagnetic bonds are elongated, and
antiferromagnetic bonds are contracted.  The ordered state is six fold
degenerate since up and down spins rows can alternate in any of the
three nearest neighbor directions and the state also has the Ising
up-down symmetry. Thus, the CTIAFM ground-state entropy density is
zero. It has been argued that there is an instability to the lattice
distortion at a temperature below $T_1$\cite{Kardar}.  In the
following section, we introduce a geometrical mapping which is valid
for all spin configurations within the ground-state ensemble of the
TIAFM and study phase transitions within this ensemble as
the coupling to the strain fields is varied. The model can be exactly
solved in this limit where excitations out of the TIAFM ground state,
the ``thermal'' fluctuations, are neglected. 
 
\section{String Picture-- CTIAFM revisited}
There is a well-known mapping of the ground states of the TIAFM on to
string configurations\cite{Blote,Dhar}. This string picture has proven
to be extremely useful in understanding the behavior of the CTIAFM.

Any spin configuration belonging to the ground state of the TIAFM has
two antiferromagnetic(satisfied) bonds and one
ferromagnetic(unsatisfied)bond in every triangular plaquette. A dimer
covering\cite{Dhar,Zeng} can be defined on the dual lattice for these
spin states. The dimers are placed on the those bonds of the dual
honeycomb lattice which connect the centers of triangles sharing an
unsatisfied (ferromagnetic) bond. Since there is exactly one
unsatisfied bond for every triangle, any site on the dual lattice is
connected to exactly one dimer. The resulting dimer covering is
unique. Fig.(\ref{dimer}a) shows such a dimer covering for an
arbitrary ground state. Under this dimer mapping, the spins in the
striped phase define a dimer covering (Fig.(\ref{dimer}b)), where all
dimers are aligned in one direction (vertical in this example). Based
on the dimer mapping, the ground states can be categorized into
different sectors by superposing a dimer covering onto the standard
dimer covering in Fig.(\ref{dimer}b) where all dimers are
vertical. The overlap defines linear structures. For example, the spin
state in Fig.(\ref{dimer}a) is mapped onto the string picture in
Fig.(\ref{string_mapping}a). Each string sector is characterized by a
string number density $p=\frac{N_s}{L}$ where $N_s$ is the total
number of strings and $L$ is system size. Under such mapping, the
disordered paramagnetic states with the maximum number of free
spins\cite{Stephenson} correspond to $p=\frac{2}{3}$ and the striped
ordered spin states correspond to $p=0$. For each string sector $p$,
the total number of spin states can be enumerated by using a
transfer-matrix technique\cite{Dhar}. The associated entropy density
is a continuous function which peaks at $p=\frac{2}{3}$ and vanishes at
$p=0$ ({\it cf} inset to Fig. \ref{free-energy}).

The strings obtained from the TIAFM ground state spin configurations
do not intersect. They run vertically through the system without being
interrupted, therefore, the number of strings per row is
conserved. Under periodic boundary conditions, strings wrap around and
form loops. If the system is restricted to the TIAFM ground state
ensemble, the number of strings does not change under any local
dynamics(single or multiple spin flips or exchanges). Thus, spin
states belonging to different string number sectors are completely
disjoint.  These string sectors are the inherent
structures\cite{Stillinger} of the TIAFM within any local dynamics:
any spin configuration will relax to one of the string sectors under
energy minimization.

Excitations out of the TIAFM ground state ensemble involve triangular
plaquettes with all three spins pointing up or down. These plaquettes
are topological defects at which the strings can
end\cite{Blote}. The TIAFM ground states and the associated dimer
coverings can be mapped to an interface model with a scalar height
variable\cite{Blote,Henley,Kondev} and the defects are
dislocations in the height field. The height mapping can be used to
demonstrate that the TIAFM ground-state is critical with power-law
correlations and the defects form the basis of the Coulomb-gas
representation of this model\cite{Blote}. Defects play an
important role in the dynamics of the strings\cite{Triest}. Single
spin flips or exchanges create defects in pairs which can then move
away from each other. Each separated defect pair is connected by two
strings. Thus, defects can increase or decrease the number of strings
by two as they move toward or away from each other.

A previous Monte Carlo study has shown that when quenched below the
first order transition temperature $T_1$ from the disordered phase, the CTIAFM 
exhibits a behavior remarkably similar to the phenomenology of
the structural glass\cite{Leigu,Triest}. For a range of supercooling
temperatures, the system remains disordered and the supercooled phase
is ergodic. Below a certain characteristic temperature $T^*$, however,
the system freezes into a ``glassy'' phase. In this phase, the system
becomes non-ergodic and the energy evolution is characterized by a
step-wise relaxation which is history dependent.

In this work, we present an analytic solution of the CTIAFM {\it
within the ground-state ensemble of the TIAFM} and show that the
coupling to the elastic strain fields leads to a phase transition at
which the string density vanishes discontinuously.  We analyze the
glassy dynamics observed in Monte Carlo simulations in terms of the
string picture and the phase transitions occurring within the
ground-state ensemble\cite{HuiPRL}.

\section{Phase transitions within the ground-state ensemble of the TIAFM}

The string picture is rigorously defined when the spin states are
restricted to the configurations without any defects.  In the TIAFM,
all these configurations have identical energies.  In the CTIAFM,
however, the energy of the state depends on the string
density\cite{HuiPRL}. To see this, one integrates out the purely
Gaussian strain fields in the CTIAFM Hamiltonian (Eq. \ref{model})
which yields an energy function with a four-spin coupling and a
coupling parameter $\mu = {\epsilon}^2{J}^2 /E$:
\begin{equation}
H = J{\sum}_{<ij>}S_i S_j
-(\mu/N){\sum}_{\alpha=1,2,3}({\sum}_{<ij>_{\alpha}}S_i S_j)^2~.
\label{four-spin}
\end{equation}
The first term is identical for all configurations in the ground state
ensemble, and can be neglected.  The sum ${\sum}_{<ij>_{\alpha}}S_i
S_j$, can be written in terms of the string density in the direction
$\alpha$, i.e., the overlap of the dimers with a configuration where
all dimers are pointing in the $\alpha$ direction. Under simple
periodic boundary conditions, where each string wraps around only once
around the system, two of the three string densities have to be equal.
In addition, the constraint that there are two good bonds per
triangular plaquette, leads to the condition that the sum of the three
string densities have to add up to the numerical value of two.  With
these constraints, only one string density is independent and the
energy per spin  of the CTIAFM can be written as
\begin{equation}
H/N =  -(\mu /2)[(1-2p)^2 +2(1-p)^2]~,
\label{string-energy}
\end{equation}
where $p$ is the one independent string density.  Since the energy
depends only on the string density and the entropy density for a given
string density has been calculated exactly\cite{Dhar}, the partition
function of the model can be calculated exactly: $Z={\sum}_{p}
\exp{(-N(\beta H(p)-\gamma (p)))} = {\sum}_p \exp{(-Nf(p))}$.  
Here $\gamma (p)$ is the entropy per spin of the string sector with
string density $p$\cite{Dhar} and $\beta$ is the inverse temperature.
The sum over $p$ can be replaced by $\exp{(-Nf_{min}(p))}$ where
$f_{min}(p)$ is $f(p)$ evaluated at the string density $p$ which
minimizes the free-energy function $f(p)$.  This free-energy function,
which is exact if excitations out of the ground-state ensemble of the
TIAFM are neglected, is shown in Fig (\ref{free-energy}).  At small
coupling constants, the free energy function shows a single global
well at $p=\frac{2}{3}$. As the coupling constant is increased, a
second minimum develops at $p=0$. A first order transition is expected
at that value of the coupling constant where the two minima become
degenerate.  In our model the four-spin coupling is of infinite range
and the barrier to nucleation is
unsurmountable. If the system is initially in
the $p=2/3$ state, it will remain indefinitely in this state.  At a
higher coupling constant, however, the $p=2/3$ state becomes unstable,
as shown in Fig. ({\ref{free-energy}). This is akin to a spinodal
point\cite{Spinodal} except that the order parameter is the string
density which involves extended structures and is not the average of
any local quantity.  The instability of the $2/3$ state is also an
entropy vanishing transition since the entropy of the $p=0$ state is
zero.

Within the ground-state ensemble of the TIAFM, no local dynamics can
change the density of strings. The entropy-vanishing transition can,
therefore, be realized in two ways; (a) allowing for the system to
explore states outside the ground-state ensemble by creating defects
and (b) implementing a non-local dynamics which can change the
string density without moving out of the ground-state ensemble.  In
this paper, we discuss the dynamics resulting from the first approach by
allowing for the presence of a small density of defects. As will be
discussed in the next section, we concentrate on analyzing the nature
of the relaxations as the entropy-vanishing transition is approached
without addressing the issue of whether this transition survives as a
true thermodynamic transition in the presence of defects.

Before proceeding to the discussion of the simulations, we would like
to point out that the the entropy-vanishing transition is akin to a
zero-temperature critical point since the ``thermal'' fluctuations, in
the form of defects, are frozen and play no role at the critical
point.  The relevant coupling is $\beta \mu$, the coupling to the elastic
strain which controls the frustration\cite{Sethna}. This scenario is
reminiscent of the zero-temperature critical point in the random-field
Ising model\cite{DFisher} with $\beta \mu$ playing the role of the random
field.

The exact solution can also be viewed in the context of inherent
structures\cite{Stillinger,Sastry} and shows that there is a phase
transition involving these structures.  At small values of $\beta
\mu$, the inherent structures belong to the $p=2/3$ sector and the
configurational entropy is finite.  As $\beta \mu$ is increased, there
is a thermodynamic instability of this sector leading to a change in
the nature of the inherent-structure landscape.  The set of inherent
structures that are accessible to the system, thus, changes at the
transition. As our simulations will show, this change can lead to an
anomalous slowing down of the dynamics.

\section{Dynamics in supercooled phase}
In this section we present a detailed discussion of the observed
dynamics in the supercooled phase.  To implement the local dynamics we
used spin-exchange kinetic extended to include updates of the
homogeneous strain fields\cite{Lei1}.  For the Monte Carlo (MC)
studies presented in this paper, a rhomboid system with periodic
boundary conditions was chosen. Unless stated explicitly, the system
size is 96x96. For all the measurements, the sampling is done every
ten MC steps. The parameters of the Hamiltonian were chosen to be;
$J=1$, ${\epsilon}=0.6$ and $E=2$. These imply a value of $\mu =0.18$
and the coupling constant $\beta \mu$ is controlled by the temperature
of the MC simulation.  In units of $\frac{J}{k_B}$ , the first-order
transition temperature is $T_1=0.667$, and the entropy-vanishing
transition temperature is $T^* = 0.397$\cite{HuiPRL}.  The defect
density at temperatures close to the entropy vanishing transition,
$T^*$, is around $0.04{\%}$\cite{HuiPRL}. We study this regime in order to
investigate the possibility of the zero-defect critical point
controlling the behavior at low but finite defect densities.

Dynamics of the supercooled phase was studied following instantaneous
quenches from a random high-temperature disordered phase into a range
of temperatures below $T_1$. After each quench, the system was
equilibrated, and the time-history of various quantities were
recorded, and auto-correlation functions were calculated.  The evidence
for diverging time scales has been presented earlier\cite{HuiPRL}.  In
this paper, we extend our analysis to the many different time scales
observed and the relationship between them.

\subsection{Relaxations of strings and spins}
The MC moves involve spin exchanges and updates of the strain
fields. Updates in strain fields change the effective interaction
between spins along the three nearest-neighbor directions and this is
reflected in the update probabilities associated with the spin
exchanges. Since the strain fields are homogeneous, these changes are
global.  

There are different classes of spins in the system.  The free spins
have equal numbers of antiferromagnetic and ferromagnetic bonds.  They
are represented as filled circles in Fig.\ref{string_mapping} and are
located at kinks in the strings. Spin exchanges involving free spins
lead to fluctuations of the strings.  Exchanges involving spins
located away from the kinks (which are not free) lead to the creation
of defects.  This is an activated process with an energy barrier of
$4J$ in the absence of any homogeneous strains.  The strain
fluctuations change this value since the effective interactions become
anisotropic and depend on the value of the strain fields.  We found
that the defect density can still be represented by an Arrhenius form
with an ``effective'' barrier which is smaller than
$4J$\cite{HuiThesis}. The two different classes of spins therefore
have very different relaxation time scales with the free spins
defining the shortest time scale in the system and the time scale
associated with the defects growing in an Arrhenius fashion.  The
dynamics of the spins is, therefore, expected to be heterogeneous and
controlled by the spatial distribution of strings.

The spatial distribution of the strings evolves in time as the strings
diffuse across the system.  Since the string density is the order
parameter associated with the entropy-vanishing transition, this is
expected to be the slowest mode in the system.  The spins, therefore,
respond to a quasi-static arrangement of the strings.  This would lead
us to expect non-exponential relaxations for the spins and the energy
fluctuations (dominated by defect number fluctuations) since
relaxation times depend on the proximity of the spins to the strings.
Our simulation results are in qualitative agreement with these
expectations.

The auto-correlation functions of the string density,
$C_{string}(t)=<p(t)p(0)>-<p(t)>^2$, at different temperatures are
shown in Fig.\ref{string_autocorr}. The
correlations functions are seen to be exponentials,
$\exp(-\frac{t}{\tau})$, with $\tau$ rapidly increasing as the
temperature decreases. The time scales $\tau$ obtained from the
fitting are plotted versus temperature in Fig.\ref{vogel}. The two
remarkable features of the string relaxation are (a) the exponential
behavior with a single time scale and (b) the rapid increase of this
time scale with an apparent divergence at a temperature close to
$T^*$.  We will discuss these features in the context of the
entropy-vanishing transition after presenting the results for the
defect, energy and spin relaxations. The rapid increase of the
string-relaxation time scales, restricts our equilibrium measurements
to temperatures $T \geq T_g \simeq 0.47$, the analog of the laboratory
glass-transition temperature for our simulations.

The auto-correlation functions of the defect number, $C_{defect}(t)=
<N_{d}(t)N_{d}(0)>-<N_{d}(t)>^2$ where $N_{d}$ is the total number of
defects, are shown in Fig.\ref{defect_corr}a.  These functions are
non-exponential and are best fit by stretched exponentials,
$\exp(-\frac{t}{{\tau}_d})^{\beta}$, with a stretching exponent
$\beta$ decreasing with temperature and a relaxation time $\tau$
increasing with temperature. The time-scale increase is Arrhenius (Fig
\ref{defect_corr}b) with no apparent finite-temperature divergence.
The stretching exponent approaches a value of $1/3$ as the temperature
approaches $T_g$.  This is consistent with a theory associating
stretched exponential relaxations with random walks on a
high-dimensional critical percolation cluster where the limiting value
of $\beta = 1/3$ is reached at percolation\cite{Almeida-Campbell}. The
energy relaxation (Fig (\ref{energy_auto})) also shows a stretched
exponential form with a $\beta$ approaching
$1/3$. Fig. (\ref{energy_auto}) also shows the waiting-time ($t_w$)
dependence of the correlations functions ({\it cf.} discussion in
section VI).  The relaxation times and stretching exponents are
summarized in Table~(\ref{defect_table}).  The energy relaxation time
shows a stronger divergence than the Arrhenius behavior of the
defects, however, the absolute values of the time scales are orders of
magnitude smaller than the string-relaxation times.  The stretched
exponential relaxation indicates that the energy and defect
relaxations are reflecting the spatial heterogeneity imposed by the
strings and their slow relaxation.

The spins are the basic microscopic entities in the system and the
relaxations of the strings and defects get reflected in the spin
relaxation. Spin auto-correlation functions at different temperatures
are shown in Fig.\ref{spin_auto}. These obey a power law decay with an
exponential cutoff; $C(t) \simeq t^\alpha \exp{(-\frac{t}{{\tau}_s})}$.
As $T_g$ is approached from above, the relaxation time ${\tau}_s$
increases exponentially and closely tracks the string relaxation
time. The exponent $\alpha$ decreases with $T$ and the value is close
to $1/4$ at $T=0.47$. The results from the fits are shown in
Table~(\ref{spin_table}). The exponent $\frac{1}{4}$ characterizes spin
relaxation in the critical ground state of the pure TIAFM. This
suggests that for times short compared to the string relaxation times,
the spins respond as they would in the TIAFM ground state ensemble,
except for the perturbation due to defect creation and
annihilation; an effect that decreases with decreasing temperature.

The results of the simulations discussed above, show that there are
multiple relaxation mechanisms in the supercooled state of the CTIAFM
and that the spin and energy relaxations become increasingly
non-exponential as the temperature approaches $T^*$.  The slowest
mode, the string-density, is characterized by a single time scale. The
string density is the `order-parameter' for the entropy-vanishing
transition and the exponential relaxation is consistent with a
mean-field transition.  Since the strings are extended objects, their
relaxations create a spatially heterogeneous environment for the local
degrees of freedom, the spins and the defects, providing a mechanism
for the stretched exponential relaxations. The observation of a
stretching exponent similar to that appearing in the theory based on
percolations clusters\cite{Almeida-Campbell} is intriguing and
suggests that the possibility of such a scenario occurring in the
CTIAFM should be explored further.

We have argued that the entropy-vanishing transition at $T^*$ can lead
to the anomalously slow dynamics observed in our simulations because
the time scale of order-parameter relaxations diverges at this
transition and because the order parameter involves extended objects
which can create spatial heterogeneities.  A remaining puzzle is the
rapidly increasing time scale associated with the string-density
relaxation. The rise in time scales is much faster than what would be
expected from usual critical slowing down.  This becomes evident upon
comparing the increase of the static susceptibility (associated with
the string density) which is expected to diverge at $T=T^*$, with the
increase in time scales.  In the temperature regime between 0.6 and
0.47, the static susceptibility increases by a factor of $2$ whereas
the relaxation time increases by a factor of $\simeq 50$. As shown in
Fig. \ref{vogel}, the rapid increase of $\tau$ can be described well
with either the Vogel-Fulcher exponential increase\cite{Vogel} or a
power law with a large exponent.  We do not have an explanation of
this anomalously fast increase of the time-scale, however, all our
observations suggest that this is an intrinsic property of the
entropy-vanishing transition and that the extended structures play a
crucial role. Further evidence supporting the claim that the entropy
vanishing transition has a different character than a usual mean-field
critical point, was provided by a study of the fluctuations in string
density over different time intervals.

\subsection{Distribution function of String Number Deviation}
The nature of the string-density fluctuations was monitored by
measuring the distribution of the string density difference
$P({\Delta}p(t))$, where ${\Delta}p=p(t+t_0)-p(t_0)$ defines the
deviation of the string density in the time $t$. The distribution is
generated by choosing different time origins $t_0$.
Fig.\ref{nongaussian} shows the distribution $P({\Delta}p)$ for
$T=0.55$ and $T=0.47$. For high temperatures, both at short and long
time intervals t, the distribution is close to a Gaussian. At $t=0$,
$P({\Delta}p)$ is just a $\delta$ function peaking at ${\Delta}p=0$.
As t increases, the width of the distribution gets broader. At some
intermediate time, the non-Gaussian behavior becomes most
prominent. After that, the distribution narrows down back to a
Gaussian and reaches a stationary Gaussian distribution.  At $T=0.55$,
the distribution become most non-Gaussian at $t=4000$. As $T$
decreases, this intermediate time scale increases rapidly, and at
$T=0.47$, the distribution becomes broader and broader with $t$, and
the stationary distribution is not observed for times as long as
$t=30,000$. According to the usual picture of the dynamics of a system
near a critical point, the distribution of the order parameter
difference is expected to become stationary at a time scale comparable
to the relaxation time which increases rapidly as the critical point
is approached\cite{goldenfeld}.  The distribution is also expected to
show significant non-Gaussian character at $T \sim T_c (L)$, where $L$
is the system size. To make a direct comparison, a measurement of the
magnetization deviation, similar to the measurement of the
string-density deviation, was undertaken for an Ising ferromagnet
on a square
lattice with $L=64$. The distribution of magnetization deviation at
time interval $t$, $P(M(t_0+t)-M(t_0))$, was  found to reach a 
stationary distribution for different $T$ as shown in
Fig.\ref{ferromagnet}. It is evident that this behavior is different
from what was observed for the strings. This difference between the
strings relaxation behavior and that of the usual order parameter,
however, is not reflected in $C_{string}(t)$. The equilibrium
$C_{string}(t)$ can be directly related to the second moment of the
distribution $P({\Delta}p(t))$ as:
\begin{eqnarray}
<({\Delta}p(t))^2> & = & <(p(t)-p(0))^2> \ =\ 2<p^2> - 2<p(t)p(0)> \nonumber \\
                   & = & 2(<p^2>-<p>^2) (1-\frac{<p(t)p(0)>-<p>^2}{<p^2>-<p>^2}) \nonumber 	
\end{eqnarray}
Thus, $$ C_{string}(t)=\frac{<p(t)p(0)>-<p>^2}{<p^2>-<p>^2} =
1-\frac{<({\Delta}p(t))^2>}{2(<p^2>-<p>^2)} ~.$$  We have measured  $<({\Delta}p(t))^2>$ and find that
$<({\Delta}p(t))^2>$ increases monotonically with $t$ despite the
non-Gaussian behavior at the intermediate time. Therefore, a
measurement of the second moment is not an adequate measure of the
complexity of the relaxation.

This picture of the string-density relaxations is sufficiently
different from the commonly accepted picture of order parameter
relaxations to justify further investigation. One obvious question
that needs to be answered is whether the defect-mediated dynamics is
responsible for the behavior or whether the zero-defect critical point
is controlling it.  We are in the process of exploring these issues.

\subsection{Spin dynamics in different string-density sectors}
To better understand the effect of ``quenched-in'' spatial
heterogeneities due to slow string relaxations, we studied spin
relaxations in different string-density sectors.
In order to fix the system in a certain string density sector,
the simulation was started from an initial spin configuration which
has the desired string density, and then was run at very low
temperature ($T=0.05$). At this temperature, the energy barrier for
creating defects is too large to be overcome within timescales
comparable to the spin relaxation times and defects are effectively
excluded from the system. String density stays at the initial value
throughout the simulation time.  The strain elastic energy scale is
much smaller than that of defect creation, and the fluctuation of
strain is finite though very small.

Fig.\ref{diff_sector} is a log-linear plot which shows the spin
auto-correlation functions for different string sectors. The nature of
the relaxation is different for different $p$. When $p \geq 0.25$, the
relaxation can be described as a power law with exponent $\simeq
0.27$. For $p \leq 0.25$, the relaxation can be best fitted to
stretched exponentials with the exponents around $\frac{1}{3}$ ({\it
cf} Table~(\ref{spin_diffSector_table})).  This analysis shows that
spin relaxations are different in different string-density sectors and
suggests that the non-exponential relaxations observed in our free
simulations, where the string density is allowed to fluctuate, is due
to this heterogeneous dynamics.

\section{Dynamics in the glass phase}

In the last section we analyzed  the dynamical behavior of the
supercooled state as it approaches $T^*$. In this section,  we look at the
dynamical behavior in the glass phase after the system is quenched
below $T^{*}$. The dynamics is studied through the measurements of
two-time correlation functions and the overlap of different copies of the
system.  We also investigate the cooling-rate dependence of various
quantities in a series of continuous cooling simulations.

\subsection{Aging}

A characteristic feature of the dynamics of many non-equilibrium
systems, including a glass, is aging. In the supercooled phase, the
system behaves as if it is in metastable equilibrium and correlation
functions are time translational invariant. In the glass phase, this
is no longer true. Although a one-time quantity such as an energy
history might show metastability as in the supercooled phase, two-time
quantities reveal that the system is evolving in an important way. The
aging of systems can be probed with the correlation functions, which
exhibit a waiting time dependence: the system behavior depends on
its age. In Fig~\ref{energy_auto}, energy auto-correlation functions
for different waiting times after the system has been quenched into
supercooled phase(T=0.55) and glass phase (T=0.45) are shown. The
aging of the system is clearly seen at T=0.45.

Under this loose definition of aging, all non-equilibrium systems
age. To distinguish between different types of aging systems,
measurements have been proposed which can classify aging systems into
different categories reflecting the complexity of the systems. We use
the approach suggested by Barrat \textit{et al}\cite{Barrat} in our
study of the CTIAFM.  These authors propose a classification
method based on the measurement of an overlap between two identical
copies of the system which distinguishes the aging of glassy systems
from domain coarsening as in an Ising ferromagnet quenched below its
critical temperature.

In order to measure the overlap, two copies of the system are prepared
with the same initial configurations and are evolved with the same
thermal noise until a time $t_w$. At time $t_w$, the two copies are
separated and subsequently evolve with different realizations of
thermal noise for a time $t$.  The function $Q(t_w,t)$ measures the
overlap of these two configurations at this time.  The quantity
$\lim_{t_w \rightarrow
\infty} {\lim_{t \rightarrow \infty} {Q(t_w,t)}}$ distinguishes different
types of aging. The limit of $t \rightarrow \infty $ can be
effectively replaced by the limit of correlation function $C(t_w,t)
\rightarrow 0$. For glassy dynamics, the overlap approaches zero as
the correlation $C(t_w ,t)$ decays to zero whereas for coarsening
systems the overlap approaches a finite value as the correlation
decays to zero\cite{Barrat}. This classification highlights the
simplicity of phase space of a coarsening system against the
complexity of phase space in a glassy system. If the phase space is
complicated, different copies of the system with the same age
continuously move apart from one another and the overlap goes to zero
in the limit of $C(t_w,t)$ going to zero. In contrast, the simple
coarsening systems  have a finite limit of this overlap. This
method has been employed in the studies of many different models
including the ferromagnetic p-spin model\cite{Swift} where this
approach was used as proof of the system being glassy.

Aging in CTIAFM was investigated by equilibrating the system in a high
temperature phase ($T=1.0$) and instantaneously quenching it to
$T=0.35$, a temperature below $T^*$.  The system size used in these
studies was 120x120. The simulation was run freely for a time $t_w$
and then three copies of the system were made and assigned different
sets of random noise. For different values of $t_w$, the correlation
function within each copy, $C(t_w,t)=\frac{1}{N}
\sum_{i}{S_i(t_w)S_i(t+t_w)}$ and the overlap between different pairs
of copies, $Q(t_w,t)=\frac{1}{N}
\sum_{i}{S_i^{(1)}(t+t_w)S_i^{(2)}(t+t_w)}$, were monitored.  The
average of these (at a given $(t_w,t)$) were stored as $Q(t_w,t)$ and
$C(t_w,t)$.  These functions are shown in the bottom panel of
Fig.\ref{overlap} and clearly demonstrate the dependence of the
correlation and overlap on the waiting time $t_w$.  The decay of these
functions become slower with increasing waiting time.  We chose to
average over regimes of $t_w$ small compared to the time over which
the correlation and overlap change significantly but large enough to
provide us with better statistics.  A more quantitative study will
have to involve better averaging of data at each $t_w$ because of the
history-dependent nature of the glass. Although $Q(t_w,t)$ and $C(t_w,t)$
decay at different rates for different $t_w$, they track each other as
can been seen from the top panel in Fig.\ref{overlap}. For values
between 0.3 and 1 of $Q(t_w,t)$ and $C(t_w,t)$, the dependence of $Q$
on $C$ is nearly linear. Below 0.3, the curves have a smaller slope
and extrapolate to zero as $C(t_w,t)$ goes to zero. The data below 0.1
(not shown in the plot) is noisy because of variations from one $t_w$
to another but definitely exhibit the trend of $Q(t_w,t)$ for
different $t_w$ approaching zero as $C(t_w,t)$ goes to zero.

The trends in overlap and correlation functions indicate that the
system is evolving in a phase space that is more complicated than that
of a simple coarsening system.  We probed this evolution at a more
microscopic level by monitoring the string density.  In
Fig.\ref{strings}(a), we show the string density as a function of time
for the master run from which copies of the system were made. The
arrows mark the different $t_w$'s at which copies were made. The
evolution of the string densities for each of the three copies,
created at a given $t_w$, are shown in
Fig.\ref{strings}(b),(c),(d). These figures demonstrate that the
overlap, $Q(t_w ,t)$ vanishes as $t\rightarrow \infty$ because the
system can explore different string sectors even when the string
density is close to zero and the waiting time is very long. The decay
rate of $Q(t_w,t)$ and $C(t_w,t)$ depends on the string sector that
the system is at, initially. The further the string density is from
$\frac{2}{3}$, the fewer the number of states available in the sector
implying stronger memory and slower decay. The tracking of $C(t_w,t)$
by the overlap $Q(t_w,t)$ is, however, an intrinsic property of the
system and does not depend on the string sector. This property implies
that as the relaxation slows down so does the rate at which two copies
meander away from each other.

In order to distinguish the overlap behavior described above from that
of a simple system at times earlier than the equilibration time, we
performed similar measurements after quenching a triangular Ising
ferromagnet with the same system size to just above $T_c$. We are
particularly interested in the overlap of the system at waiting times
smaller than the equilibration time.  As seen from
Fig.~\ref{ferro_overlap}, at a waiting time $t_w=50$, the overlap
approaches a finite value as the correlation decays to zero. At longer
$t_w$, after the system has reached equilibrium, the overlap and
correlation function are independent of $t_w$.  In this regime, it can
be easily shown that the overlap is trivially related to the
correlation function and always goes to zero following the
correlation\cite{overlap_at_equil}. The short waiting time behavior of
the ferromagnetic model is obviously different from what we saw in
CTIAFM and we would like to attribute this difference in behavior to
the difference in the complexity of the free energy landscape.

\subsection{Effects of cooling rate}
The easiest way to obtain a glass from a liquid is to cool the liquid
fast enough. If the relaxation time scale of the liquid at a certain
temperature becomes larger than the time scale associated with the
cooling rate, the liquid fails to reach equilibrium and becomes a
glass. Different cooling rates  cause the liquid to fall out of
equilibrium at different temperatures, which implies different
laboratory glass transition temperatures. The resulting glass is a
non-equilibrium system and its properties will in general depend on its
history of production. In this section we will explore the cooling
rate dependence of CTIAFM. In Monte Carlo simulation of cooling the
model glass, we define the cooling rate ($\gamma$) as $$\gamma =
\frac{dT}{dt}\ ,$$ where $dt$ is the number of Monte Carlo steps per spin over which the  
temperature changes by $dT$. The simulations were started with the
equilibrium configuration at high temperature T=5.0, and the energy
was measured as a function of temperature during the cooling run. The
temperature dependence of the energy for different cooling rates is
shown in Fig.~\ref{cooling}. As seen from Fig.~\ref{cooling}, the
faster cooling rates make the system fall out of equilibrium at higher
temperatures and the energy at the end point is higher. A closer look
at the end configurations has shown that for different cooling rates,
the end configurations at T=0 all belong to the $\frac{2}{3}$ string
density sector, but with different defect densities. The faster the
cooling rate, the higher the defect density and no local ordering was
observed at the end of the cooling runs for any of the cooling
rates. Since this is a mean field model and local strain fluctuations
are not allowed, such local ordering is suppressed. With the time
scales corresponding to the cooling rates, the string density does not
have enough relaxation time to explore string density sectors other
than $\frac{2}{3}$. At low temperatures, where the defect density is
small, the energy of the system can be written as $E(T) \simeq
E_{string}(p)+E_{defect}(T)$ where the first term is the energy of
string sector $p$ in zero defect situation, and the second term is the
excitation energy arising from a non-zero defect density, $\rho (T)$,
and can be written as $E_{defect}(T)= E_0 {\rho}(T)$. For infinitely
slow cooling, the defect number density ${\rho}(T)$ is expected to be
Arrhenius. Since the system gets stuck at $p=\frac{2}{3}$,
$E_{string}(p)$ is a constant equal to $E_{string}(\frac{2}{3})$. The
energy fluctuation of the system is mainly from the contribution of
$E_{defect}(T)$, the dynamics of the system is dominated by the
relaxation of defects.  In this picture, the system will fall out of
equilibrium at temperatures where ${\rho}(T)$ falls out of equilibrium
at different cooling rates. So by cooling continuously into low
temperature regime we are essentially probing the dynamical behavior
of the defects.  From the measured dependence of the energy on
temperature, we have extracted the behavior of $\rho (T)$ and compared
it to the {\it equilibrium} defect density which has an Arrhenius
form\cite{HuiThesis}. As can be seen from Fig.~\ref{defect-cooling},
the defect number density curve ${\rho}(T)$ deviates from the
equilibrium curve, and the deviation occurs at lower temperatures for
lower cooling rates.  We, therefore, conclude that the cooling rate
dependence of the energy in the CTIAFM arises from the freezing in of
non-equilibrium defect densities. If the CTIAFM was subjected to a
steepest descent minimization of energy at the end point
configurations obtained from the cooling runs, the energy would be
that of the $p=2/3$ state and this is the limiting value reached for
arbitrarily small cooling rates.  In this sense, the $p=2/3$
configurations with no defects frozen in is the ideal glass that would
be obtained from slow cooling.

\section{Conclusion}

In this paper, we have presented a detailed study of a non-randomly
frustrated spin system which exhibits glassy behavior as exemplified
by non-exponential relaxations, rapidly diverging time scales and
aging.  The crucial features of the model which were related to the
glassy dynamics are (a) the presence of extended spatial structures
and (b) a phase transition involving these structures which is driven
by a parameter controlling the frustration in the system.  The
extended spatial structures are reminiscent of the dynamical
heterogeneities observed in experiments\cite{Weitz} and simulations of
Lennard-Jones liquids\cite{Glotzer}.  One of the conjectures based on
our study, is that these dynamical heterogeneities are a consequence
of the frustration in the system and they are made up of particles
which are in the most energetically unfavorable positions.  This
conjecture should be experimentally verifiable.

In our model, we have argued that the presence of a thermodynamic
phase transition is responsible for the glassy behavior.  The nature
of this phase transition is unusual in that the time scale divergence
is much stronger than what would be expected based on the dynamics of
usual thermal critical points.  We have no clear understanding of the
source of this unusual behavior, however, it seems certain that the
presence of the extended structures is a crucial factor.  This observation
leads to the intriguing possibility that a similar transition,
involving the dynamical heterogeneities, underlies the glassy behavior
in supercooled liquids.  Measurement of correlation functions related
to the dynamical heterogeneities should shed some light on this issue.

Further work is now in progress to identify the exact nature of the
phase transition in our model.  The main question that we are
addressing is the reason for the rapid divergence of time scales. The
scenario we have observed is reminiscent of the transition in
random-field Ising models\cite{DFisher} and understanding this
similarity should go a long way toward answering the question of what
plays the role of the quenched randomness in a supercooled liquid.

The work of BC was supported in part by NSF grant number DMR-9815986
and the work of HY was supported by DOE grant DE-FG02-ER45495. We
would like to thank R. K. Zia, W. Klein, H. Gould, S. R. Nagel and
J. Kondev for many helpful discussions.

\newpage
\vspace{0.2in}
\begin{figure}[h]
\epsfxsize=4.5in \epsfysize=2in
\epsfbox{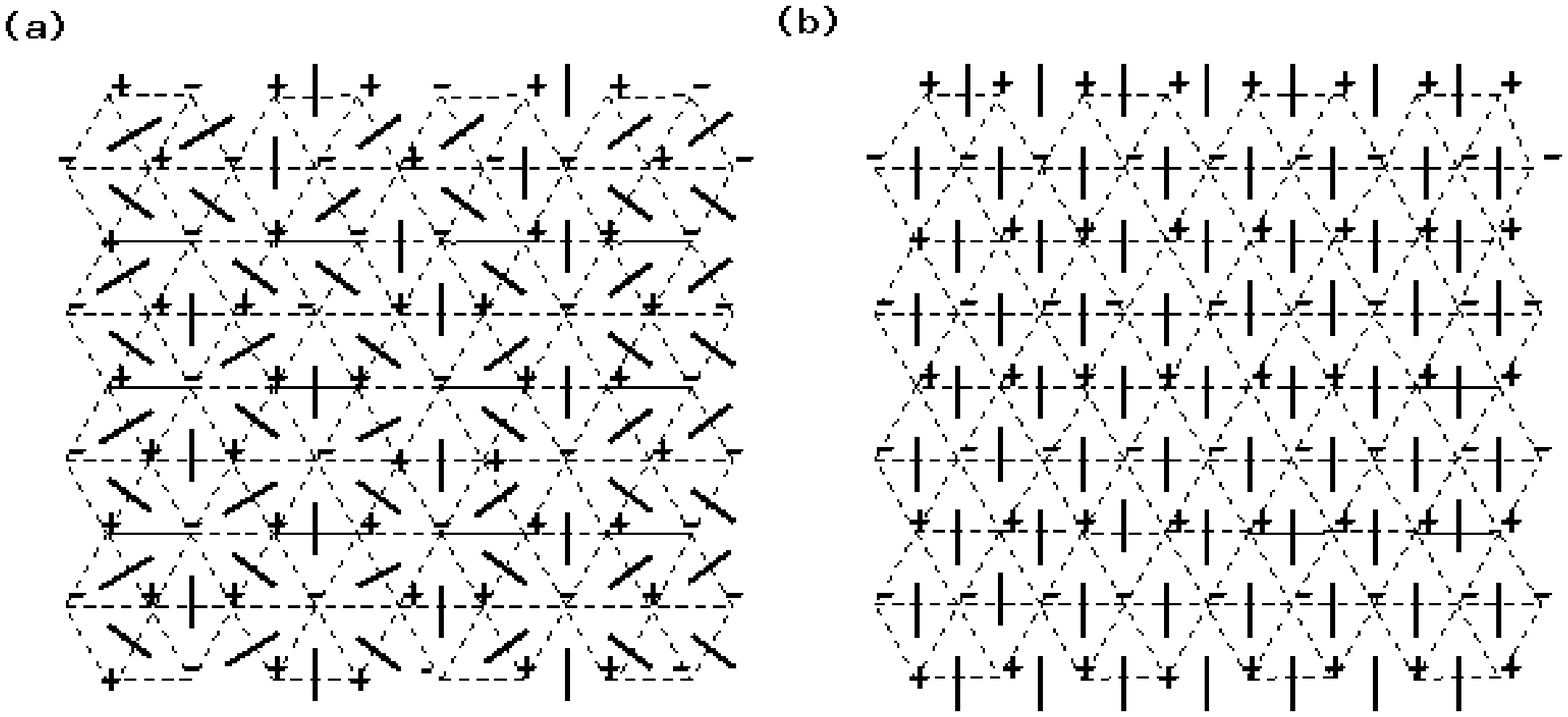}
\vspace{0.2in}
\caption{A dimer covering corresponding to a spin configuration can be obtained by
putting dimers crossing the ferromagnetic bond connecting the centers
of the two triangles sharing this bond. (a)the dimer covering for a
random TIAFM ground state configuration (b) A special dimer covering
which is mapped from the striped-order spin configuration and
characterized by all vertical dimers. This dimer covering serves as a
reference in defining the string picture.}
\label{dimer}
\end{figure}
\vspace{0.2in}
\begin{figure}[h]
\epsfxsize=4.5in \epsfysize=2in
\epsfbox{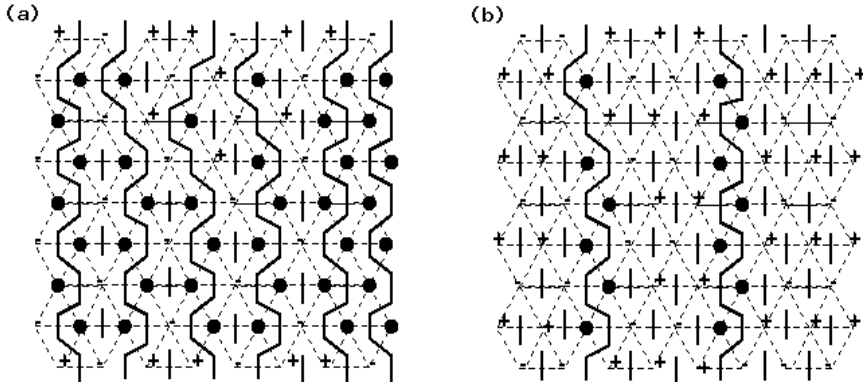}
\vspace{0.2in}
\caption{A string picture is obtained by overlapping a dimer covering with the
reference dimer  covering Fig.\ref{dimer}b. The filled circles are free spins,
the spins which have 3  satisfied and 3 unsatisfied bonds. As can be seen, the free spins are
always tied to the strings.  (a) the string mapping from the spin configuration
Fig.\ref{dimer}a (b) a  string mapping which has a fewer number of strings.}  
\label{string_mapping}
\end{figure}
\newpage
\begin{figure}[h]
\epsfxsize=4.5in \epsfysize=4.5in
\hspace{0.3in}
\epsfbox{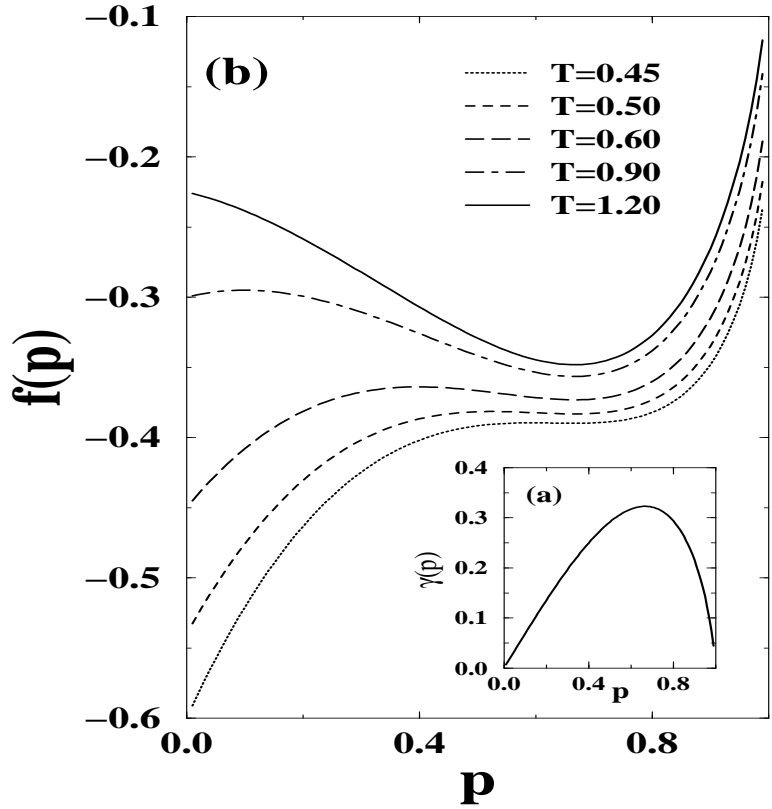}
\caption[Entropy and the dimensionless free energy as functions of the string 
density $p$]{(a) Entropy as a function of the string density from the
work of Dhar {\t et al.}~\cite{Dhar} (b)The dimensionless free energy
$f(p)$ for $\mu =0.18$. $T^*$ for this value of $\mu$ is 0.397.
Temperature is measured in units of $1/k_{B}$.}
\label{free-energy}
\end{figure}
\newpage
\begin{figure}[h]
\epsfxsize=4.5in \epsfysize=4.5in
\hspace{0.6in}
\epsfbox{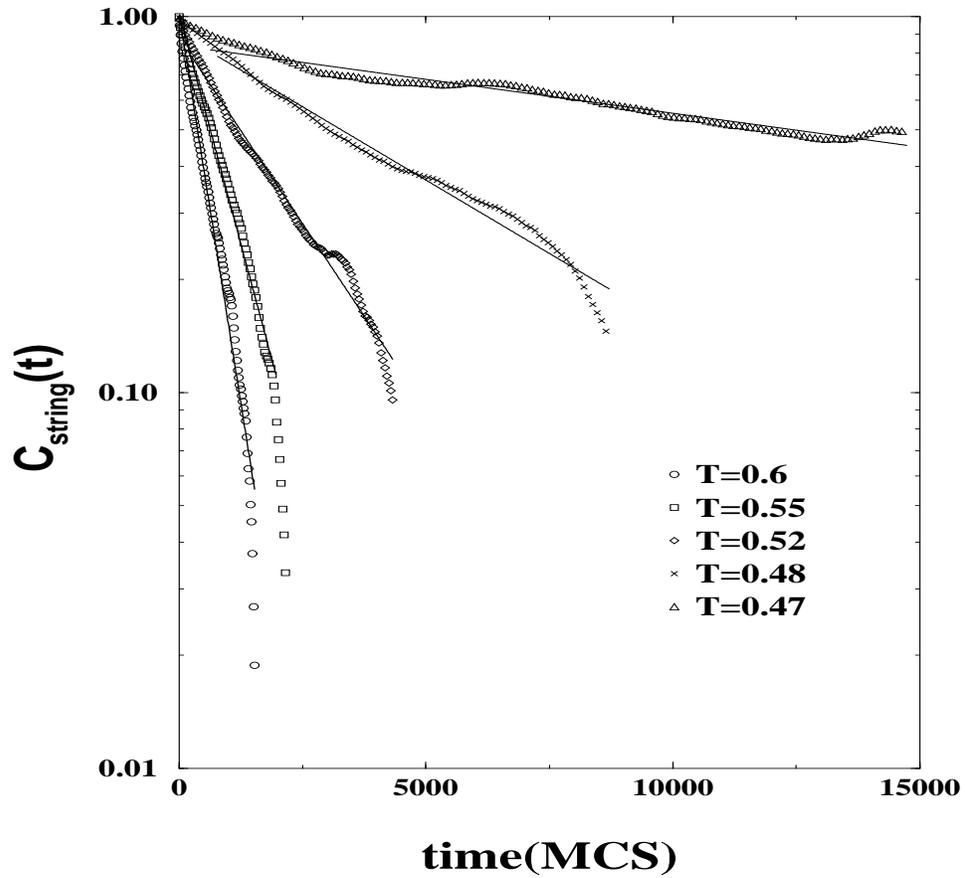}
\vspace{0.3in}
\caption{Auto-correlation functions of string density at different
temperatures. The curves have been fitted to exponentials
$\exp(-\frac{t}{\tau})$.}
\label{string_autocorr}
\end{figure} 
\newpage
\begin{figure}[h]
\epsfxsize=4.5in \epsfysize=4.5in
\hspace{0.5in}
\epsfbox{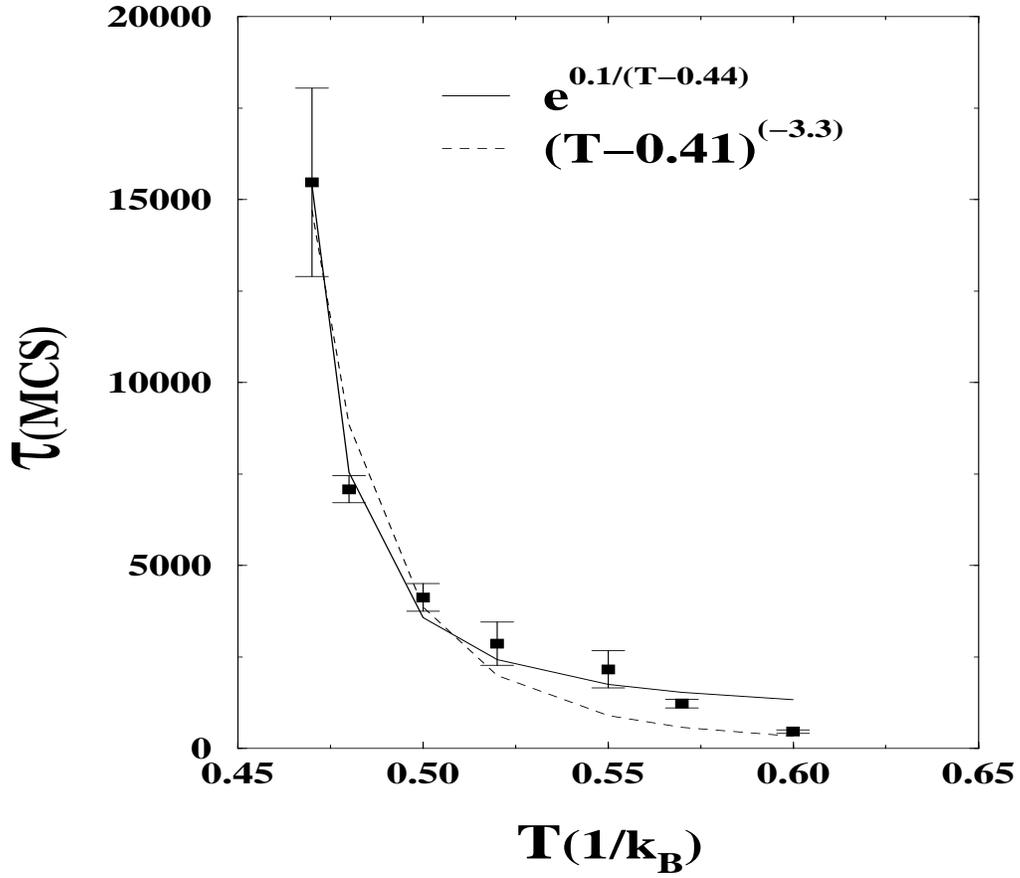}
\vspace{0.4in}
\caption{String density correlation time $\tau$, extracted from the fitting 
in Fig.~\ref{string_autocorr}, shown as a function of
temperature. Fits to a Vogel-fulcher law (solid line) and a power law
(dashed line) are also shown.}
\label{vogel}
\end{figure}
\newpage
\begin{figure}[h]
\epsfxsize=5.5in \epsfysize=5.5in
\hspace{0.4in}
\epsfbox{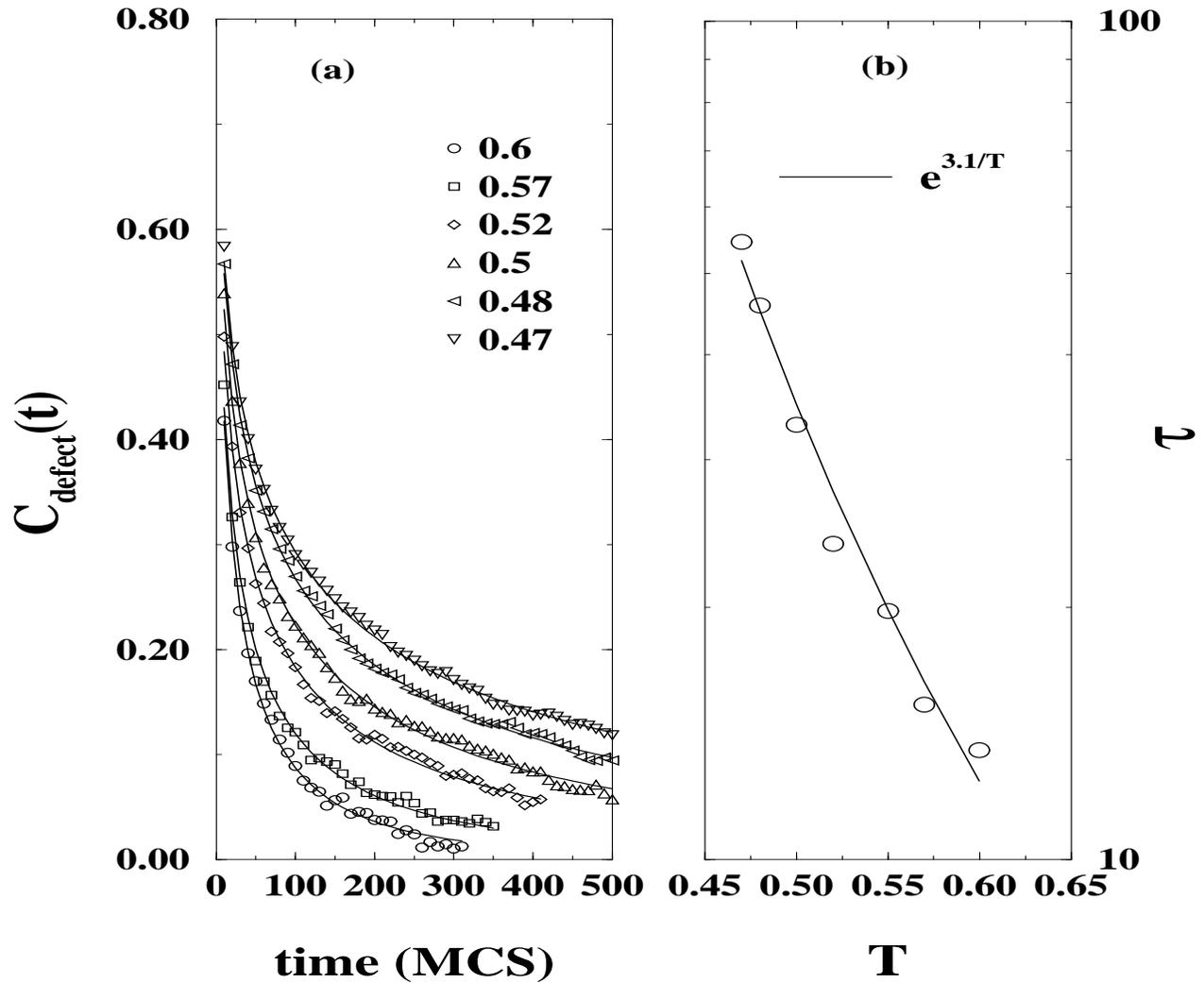}
\vspace{0.3in}
\caption{Defect number auto-correlation functions and correlation times at
different temperatures. (a) The auto-correlation functions can be
fitted to stretched exponentials
$\exp(-\frac{t}{{\tau}_d})^{\beta}$. The stretching exponent $\beta$
decrease with $T$ and (b) ${\tau}_d$ increases with $T$ in an Arrhenius
fashion.}
\label{defect_corr}
\end{figure}
\newpage
\begin{figure}[h]
\epsfxsize=4.5in \epsfysize=4.5in
\hspace{0.3in}
\epsfbox{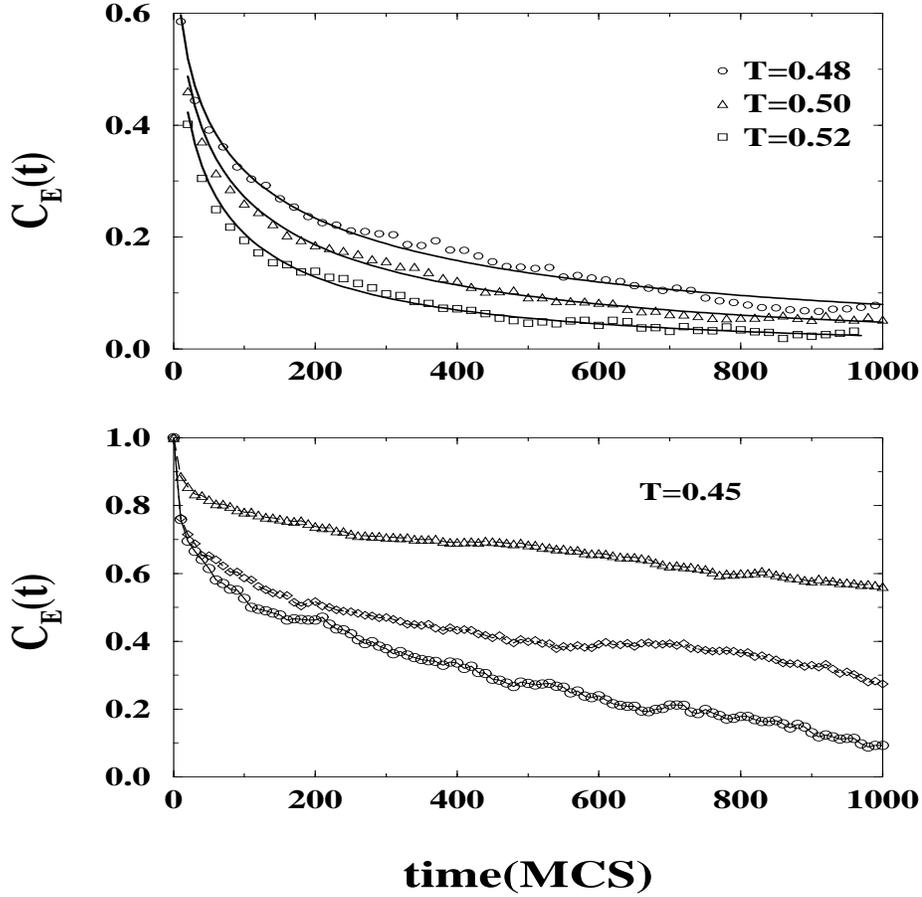}
\vspace{0.4in}
\caption{The energy auto-correlation functions, $C_E(t_w,t)$, at different 
temperatures
above $T_g$(top panel) and at different waiting times at a temperature
$T=0.45$ which is below $T_g$ (bottom panel). In the top panel, the
solid lines are the stretched exponential fits.   The different curves at
$T=0.45$ are obtained from the measurement of $C_E(t)$ over different
ranges of $t_w$. From bottom to top, these ranges are $0<t_0<25000,
18000<t_0<48000$ and $50000<t_0<80000$.}
\label{energy_auto}
\end{figure}
\newpage
\begin{figure}[h]
\epsfxsize=4.5in \epsfysize=4.5in
\hspace{0.3in}
\epsfbox{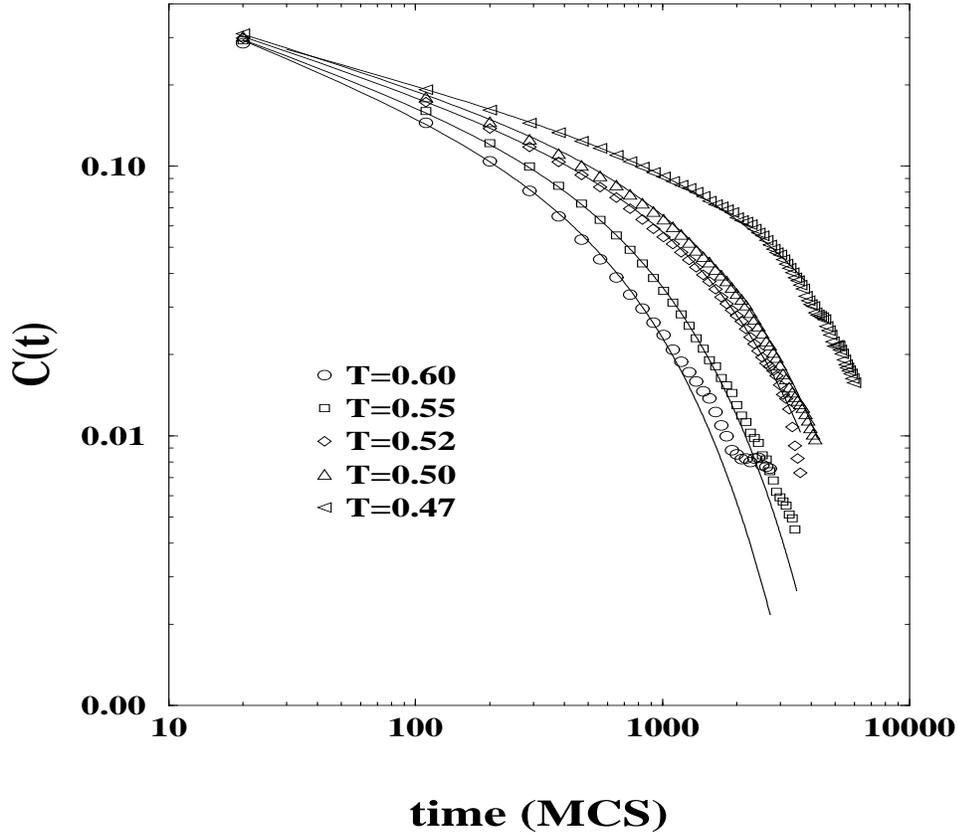}
\vspace{0.4in}
\caption{Spin auto-correlation functions at different temperatures above
$T_g$. The curves have been fitted to power law decays with an
exponential cutoff, $t^\alpha \exp(-\frac{t}{{\tau}_s})$.}
\label{spin_auto}
\end{figure}
\newpage
\begin{figure}[h]
\epsfxsize=4.5in \epsfysize=4.5in
\hspace{0.1in}
\epsfbox{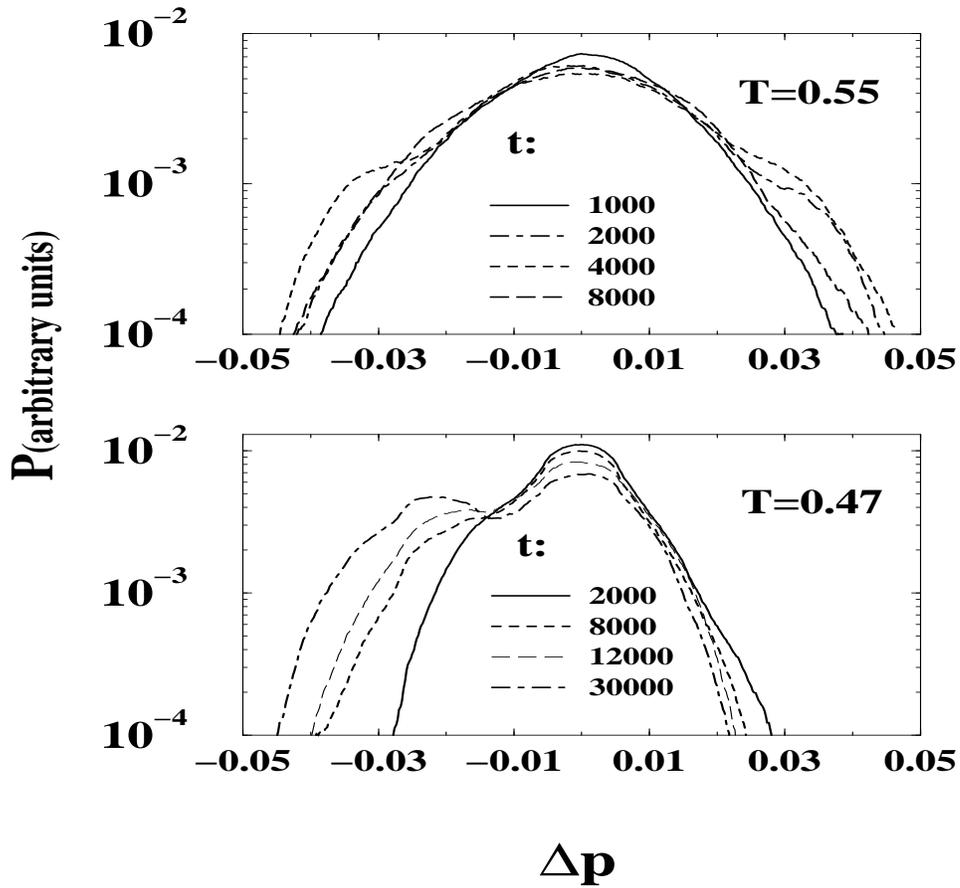}
\vspace{0.2in}
\caption{The distribution functions of the string density deviation at various
time differences $t$.  The distribution is generated by choosing
different time origins $t_0$. The areas under the curves have been
normalized to unity. At T=0.55, a prominent non-Gaussian behavior of
the distribution function is seen at $t=4000$, after which the
distribution narrows back down to Gaussian. At T=0.47, the
distribution becomes broader and broader with $t$, and the stationary
distribution is not observed for times as long as 30,000.}
\label{nongaussian}
\end{figure}
\newpage
\begin{figure}[h]
\epsfxsize=3.5in \epsfysize=3.25in
\hspace{0.8in}
\epsfbox{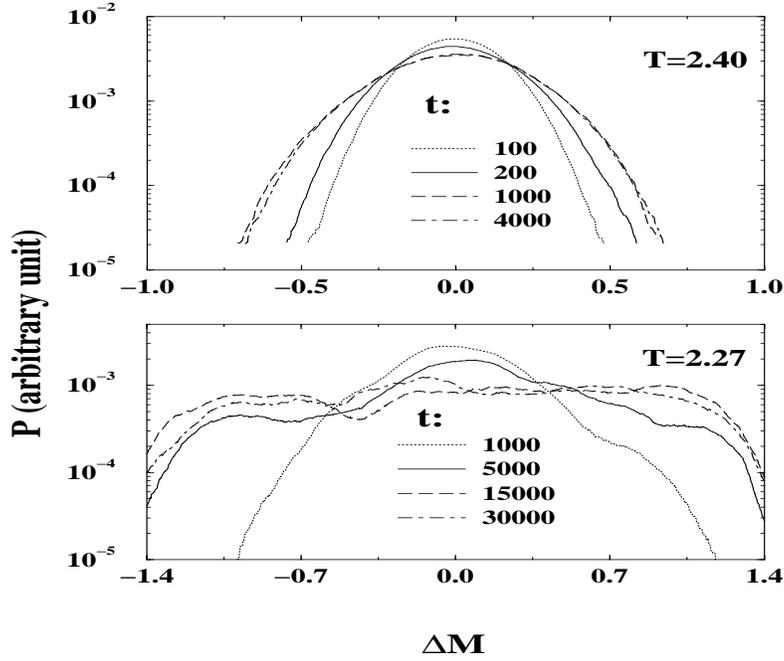}
\vspace{0.4in}
\caption{The distribution functions of the magnetization deviation at various
time difference $t$ measured in a square lattice Ising ferromagnet
with system size $L=64$. The critical temperature $T_c(L) \simeq
2.27$. The areas under the curves have been normalized to unity. The
relaxation behavior of the order parameter at the ferromagnetic critical point
is seen to be different from that of the strings in CTIAFM.}
\label{ferromagnet}
\end{figure}
\newpage
\begin{figure}[h]
\epsfxsize=4.5in \epsfysize=4.5in
\hspace{0.3in}
\epsfbox{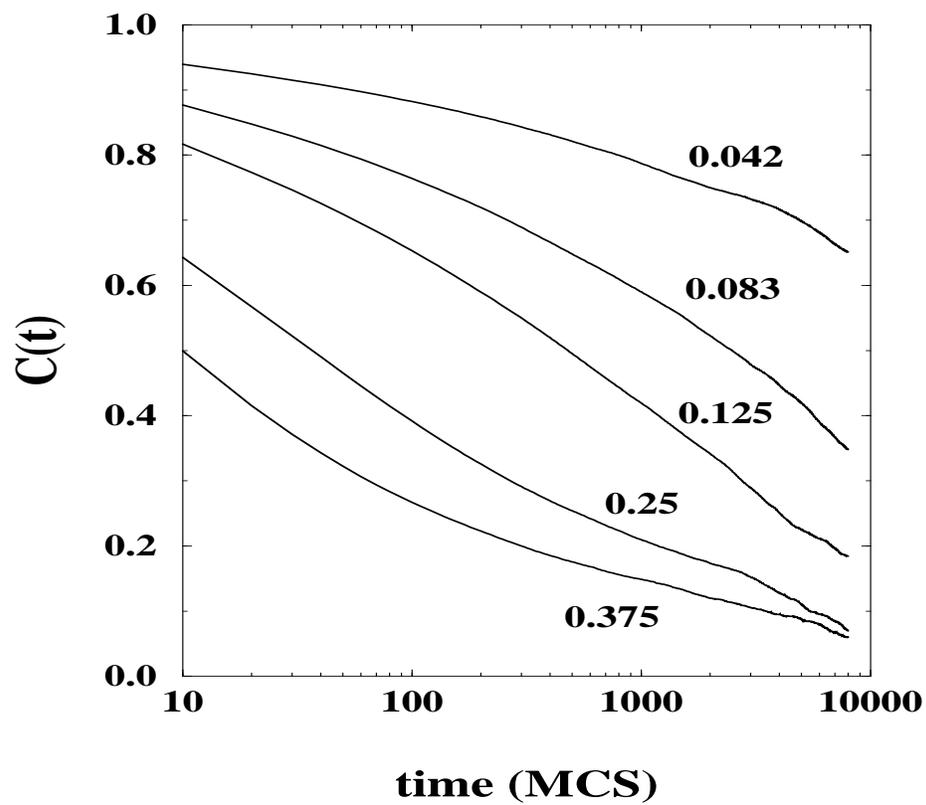}
\vspace{0.2in}
\caption{Spin auto-correlation functions for different string density
sectors,  $p$. For $p \geq 0.25$, the relaxation can be described by a
power law.  For $p \leq 0.25$, the
relaxation is  best fit to a
stretched exponential form.}
\label{diff_sector}
\end{figure}
\newpage
\begin{figure}[h]
\epsfxsize=4.5in \epsfysize=4.5in
\hspace{0.3in}
\epsfbox{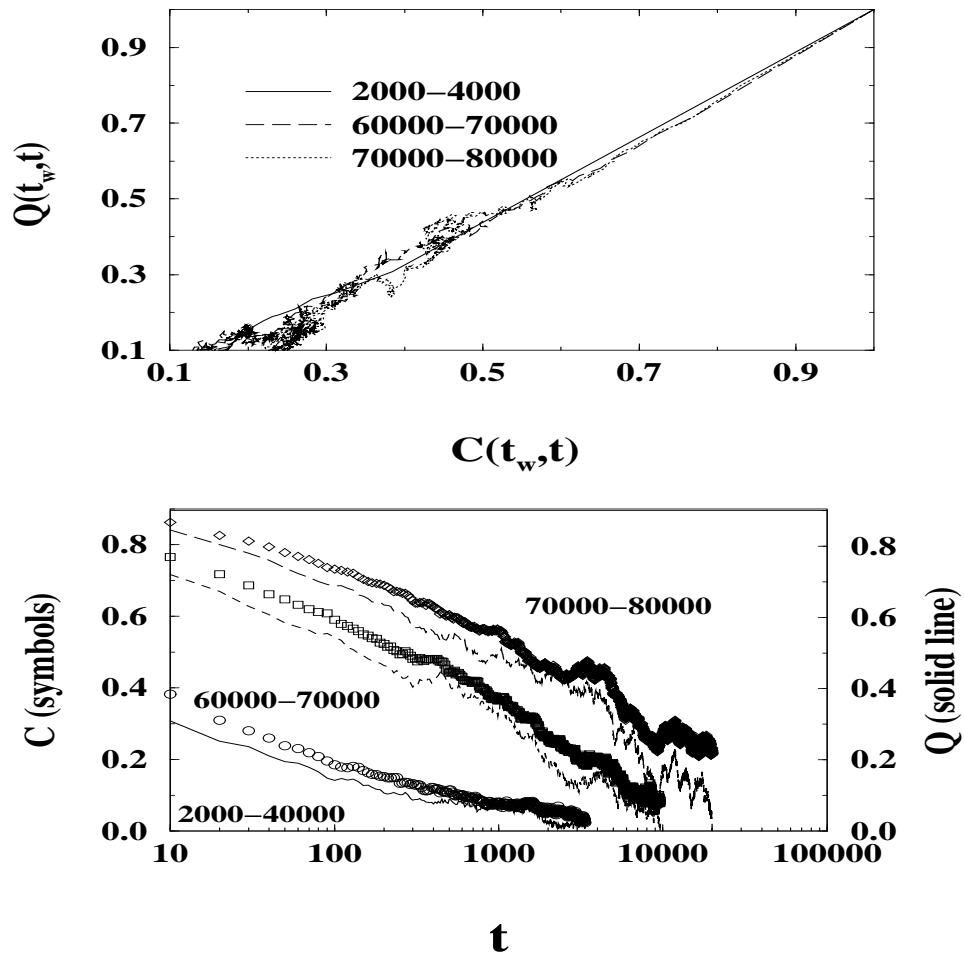}
\vspace{0.5in}
\caption{The top panel shows the overlap,
$Q(t_w,t)$, versus the correlation function $C(t_w,t)$.  In the bottom
panel $C(t_w,t)$ and $Q(t_w,t)$ are shown for different ranges of
$t_w$.}
\label{overlap}
\end{figure}
\newpage
\begin{figure}[h]
\epsfxsize=4.5in \epsfysize=4.5in
\hspace{0.3in}
\epsfbox{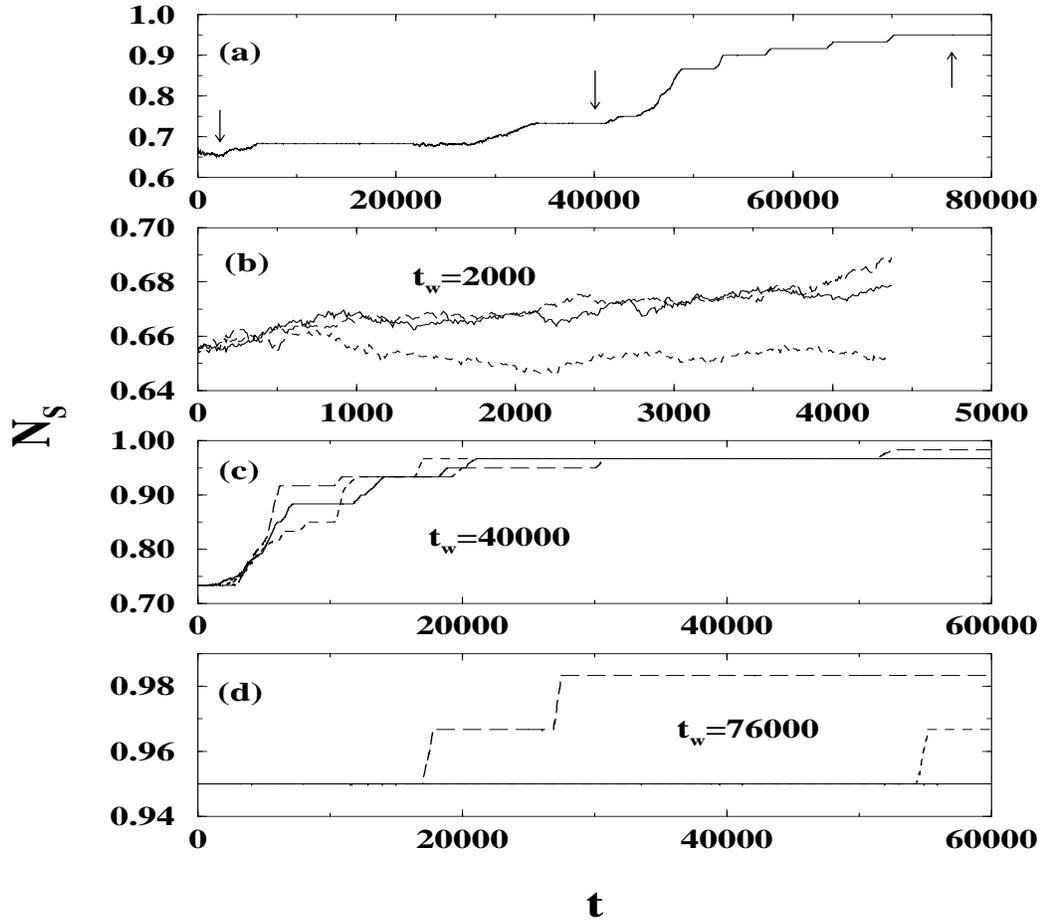}
\vspace{0.6in}
\caption{(a)Time history of the string density after the CTIAFM is quenched
to $T=0.35$. At each of the different times marked by arrows in (a),
three copies of the system are made and evolved with independent noise
realizations. The panels (b),(c) and (d) depict the history of string
density of the three copies made at $t_w=2000, 40000,$ and $76000$,
respectively.}
\label{strings}
\end{figure}
\newpage
\begin{figure}[h]
\epsfxsize=4.5in \epsfysize=4.5in
\hspace{0.3in}
\epsfbox{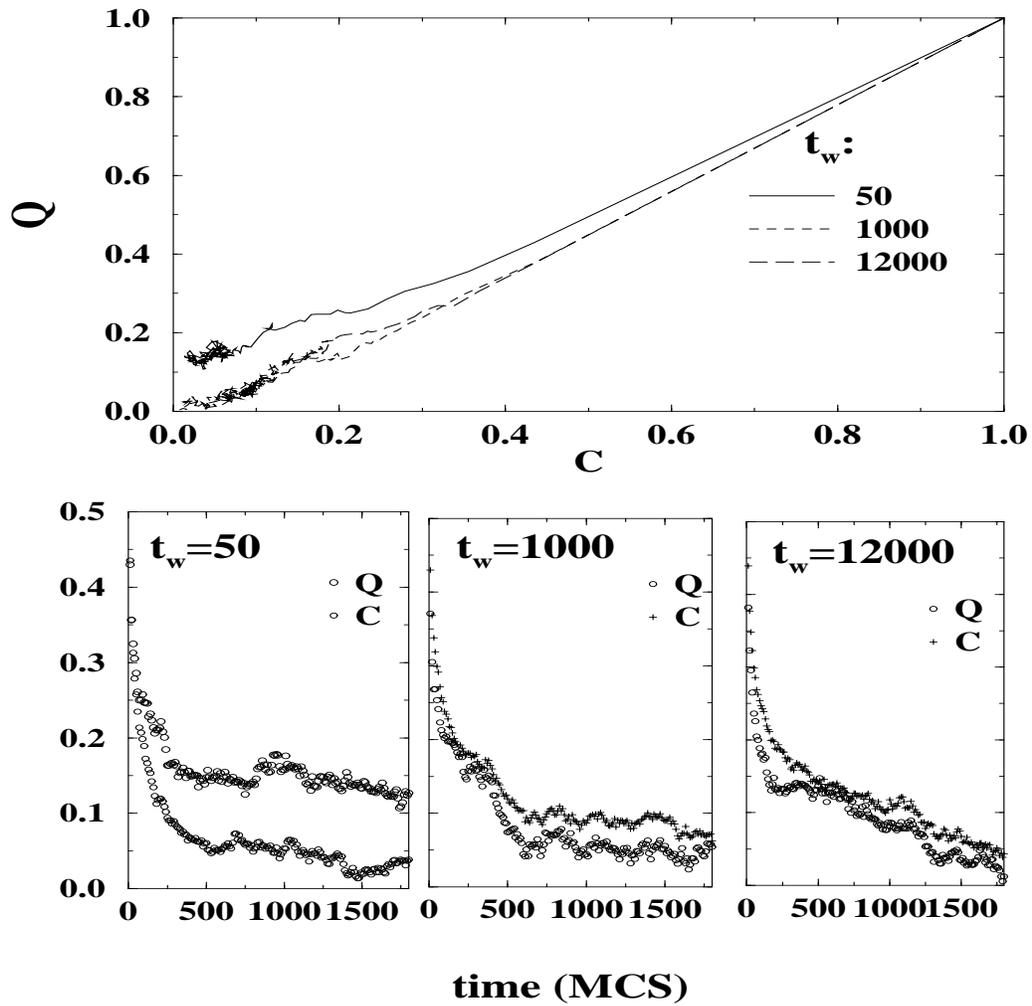}
\vspace{0.6in}
\caption{$Q(t_w,t)$ and $C(t_w,t)$ after quenching triangular ferromagnet to
just above $T_c$. From the top panel, $Q(t_w,t)$ is seen to reach a
finite value for $t_w=50$ and vanish for $t_w=1000$ and $12000$ as
$C(t_w,t)$ decays to zero. In bottom three panels $Q(t_w,t)$ and
$C(t_w,t)$ are shown as functions of time for different $t_w$.  The
behavior of $Q(t_w,t)$ at small $t_w$ when the system is yet to reach
equilibrium is seen to be different from that of the CTIAFM in the
glass phase.}
\label{ferro_overlap}
\end{figure}
\newpage
\begin{figure}[h]
\epsfxsize=4.5in \epsfysize=4.5in
\hspace{0.3in}
\epsfbox{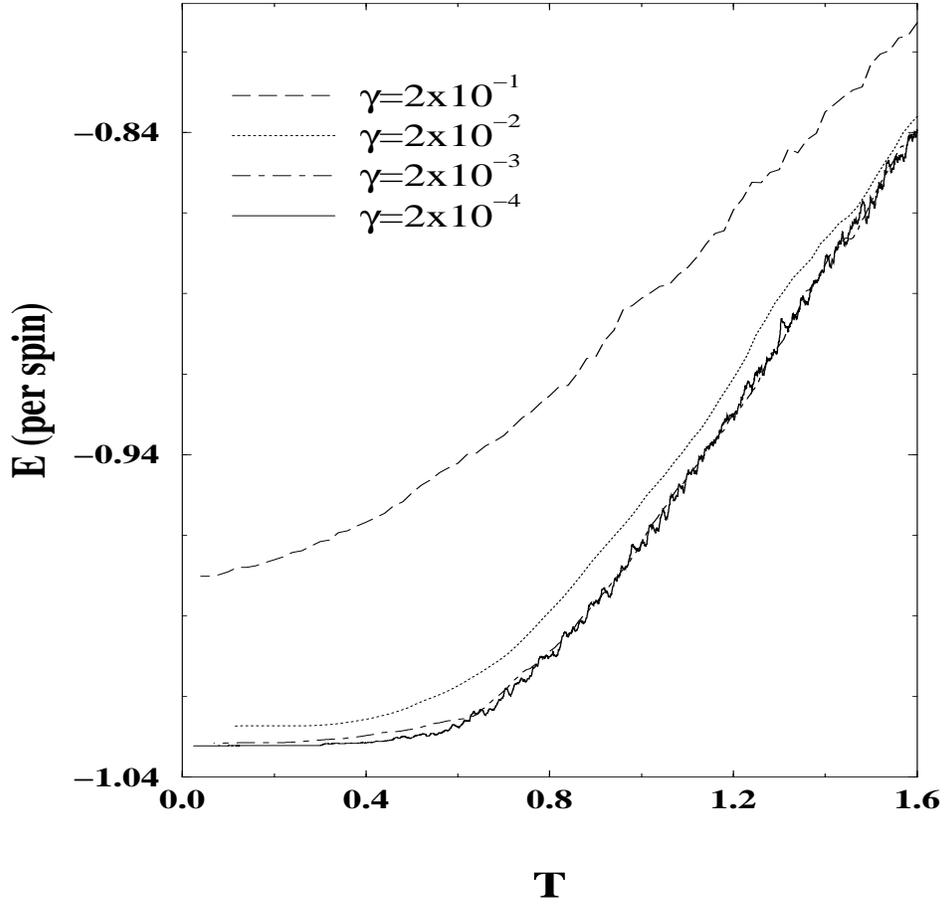}
\vspace{0.4in}
\caption{The temperature dependence of the energy $E$ for  different cooling rates $\gamma$.}  
\label{cooling}
\end{figure}
\newpage
\begin{figure}[h]
\epsfxsize=4.5in \epsfysize=4.5in
\hspace{0.3in}
\epsfbox{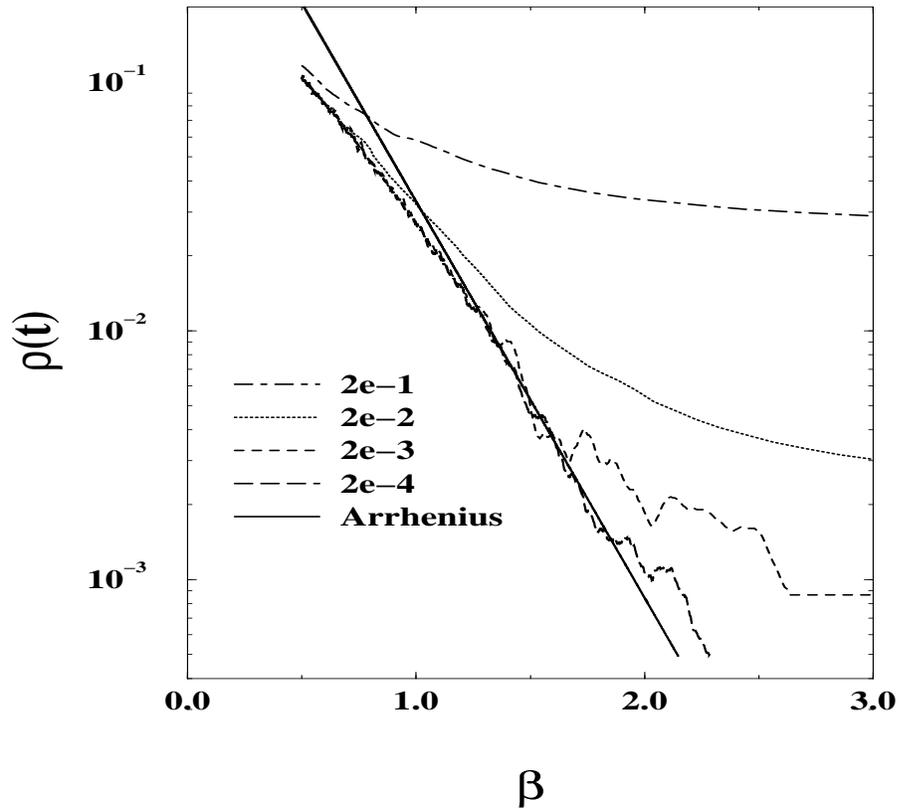}
\vspace{0.3in}
\caption{Defect density $\rho(t)$ as a function of $\beta=T^{-1}$ for different  cooling rates, $\gamma$, compared to the equilibrium, Arrhenius form.}
\label{defect-cooling}
\end{figure}
\newpage
\begin{table}
\caption{Results from fitting the string auto-correlation functions and defect number auto-correlation functions. The first column is the temperature. The second and third are the correlation time of the string density and its associated error bars from the exponential fit. The next two columns are the values of $\beta$ and ${\tau}_d$ from the defect number auto-correlation functions fitted to a stretched exponential form. The last two columns are the values of ${\tau}_E$ and $\beta$ from the energy auto-correlation fitted to a stretched exponential form.}
\vspace{0.2in}
\label{defect_table}
\footnotesize
\begin{tabular}{|c|c|c|c|c|c|c|c|c|} \hline
$T$ & $\tau\ (string)$ & $\Delta$$\tau$\ (string)&  & $\beta\ (defect)$ & $\tau_{d}\ (defect)$ & & $\tau_{E}\ (Energy)$ & $\beta\ (Energy)$ \\ \hline
0.6 &  457  & 45  &  & 0.44 $\pm$ 0.01  & 13.5 & & 12.6 & 0.42 $\pm$ 0.004  \\ 
0.57 & 1223 & 119 &  & 0.39 $\pm$ 0.01  & 15.3 & & 13.4 & 0.38 $\pm$ 0.01 \\
0.55 & 2161 & 505 &  & 0.38 $\pm$ 0.01  & 19.8 & & 23.0 & 0.40 $\pm$ 0.04 \\ 
0.52 & 2860 & 593 &  & 0.36 $\pm$ 0.02 & 23.8 & & 29.8 & 0.38 $\pm$ 0.02 \\ 
0.50 & 4124 & 378 &  & 0.37 $\pm$ 0.02  & 33.0 & & 49.0 & 0.35 $\pm$ 0.01 \\ 
0.48 & 7081 & 370 &  & 0.34 $\pm$ 0.02  & 45.8 & & 67.9 & 0.34 $\pm$ 0.03 \\ 
0.47 & 15472& 2575&  & 0.33 $\pm$ 0.01 &  54.5 & & 87.2 & 0.32 $\pm$ 0.03 \\ \hline
\end{tabular} 
\normalsize
\end{table}
\begin{table}
\caption{The correlation time $\tau_{s}$ and exponent $\alpha$ extracted from the fitting of the spin auto-correlation function to the form $C\ t^{-\alpha}\exp(-\frac{t}{{\tau}_s})$ at different temperatures.}
\vspace{0.2in}
\label{spin_table}
\begin{tabular}{|c|c|c|c|} \hline
$T$ & ${\tau}_s\ (spin)$ & $\alpha\ (spin)$ & C (prefactor) \\ \hline
0.6 &  865 & 0.36  & 0.88 \\ 
0.55 & 1140 & 0.32 & 0.78 \\ 
0.52 & 1980 & 0.30 & 0.75 \\ 
0.50 & 2170 & 0.28 & 0.72 \\ 
0.47 & 4910 & 0.25 & 0.64 \\ \hline
\end{tabular} 
\end{table}
\begin{table}
\caption{Results obtained from fitting  the spin auto-correlation functions 
in different string-density sectors to the stretched exponential form
$\exp{(-(t/\tau_{s})^{\beta})}$}.
\vspace{0.3in}
\label{spin_diffSector_table}
\begin{tabular}{|c|c|c|} \hline
$string\ density$ & ${\tau}_s\ (spin)$ & $\beta\ (spin)$\\ \hline
p=0.042 & $1.9 \times 10^5$ & 0.28  \\ 
p=0.083 & 7435 & 0.32  \\ 
p=0.125 & 1522 & 0.33 \\ \hline
\end{tabular} 
\end{table}


\begin{thebibliography}{99}
\bibitem{overlap_at_equil} It can be shown that for systems in equilibrium, 
$Q(t)$, the overlap between two copies evolving
independently for time $t$, is equal to $C(2t)$~\cite{Barrat}.
\end{thebibliography}
\end{document}